\documentclass[11pt]{article}
\usepackage{amsmath,amssymb,amsthm}

\newtheorem{theorem}{Theorem}

\newtheorem{proposition}[theorem]{Proposition}

\advance \topmargin by -\headheight
\advance \topmargin by -\headsep
\evensidemargin \oddsidemargin
\title{An Introduction To Regret Minimization In Algorithmic Trading: A Survey of Universal Portfolio Techniques}
\author{Thomas Orton}
\date{April 2017}

\begin{document}

\maketitle

\section{Introduction}

Consider the following problem: you have $\$1$ which you can invest in a stock market with $m$ stocks, and you would like to invest this money over $T$ trading periods. How should you invest your money so that the amount of wealth you end up with isn't too far away from the wealth you could have achieved had you been able to see the future prices of all stocks \textbf{before} you made your investment decision? 

\subsection{Notation}

To make this question more precise, let's introduce some notation. For $i \in [1,T]$, let $x_i \in \mathbb{R}^m_{+}$ denote the \textbf{price relative vector} between days $i-1$ and $i$. In particular, this means that $(x_i)_{j}:=\frac{\text{price of stock $j$ at time $i$}}{\text{price of stock $j$ at time $(i-1)$}}$, and we assume that $\forall i,j, (x_i)_{j} > 0$. For $i \in [1,T]$, let $b_i \in \Delta^m$ denote the \textbf{portfolio vector} at time $i$, where $\Delta^m$ is the $(m-1)$ dimensional simplex. This means that $(b_i)_{j}$ is the proportion of one's current wealth at time $i$ which is invested in stock $j$. For example, suppose that at time $i \in [1,T]$ we have already traded for $(i-1)$ time periods, and as a result our current earnings is $S$. At time $i$, we pick a portfolio vector $b_{i}$, and $x_{i}$ is then revealed to us. Our new earnings become $S \langle b_{i}, x_i \rangle$. Notice that we assume our entire wealth is reinvested into the stock market at each new time period. This is without loss of generality, since we can always fabricate a constant "savings" stock which never changes in price (investing in this stock would be equivalent to not investing in the market). Now, if we start with $\$1$, this means that our earnings after $T$ time periods is $S_{T}:=\prod_{k=1}^T \langle b_i,x_i \rangle$. 

\subsection{Constant Rebalanced Portfolios (CRPs)}

How should we measure the performance of $S_{T}$ on a stock market? Ideally, we would like to make minimal assumptions on the sequence $\{x_i\}_{i=1}^T$. An approach which has largely been adopted is to consider a class of experts of Constant Rebalanced Portfolios, and then measure the difference between $S_{T}$ and the earnings the best-in-class expert achieved. A CRP strategy is one which keeps a fixed proportion of ones wealth in each stock for every time period. Using the notation we have introduced, a CRP picks a single $b \in \Delta^m$ and achieves earnings $S_{T}(b)=\prod_{k=1}^T \langle b,x_i \rangle$. For example, imagine a stock market with two stocks, where $x_{i}=(\frac{1}{2},1)$ for odd $i$ and $x_{i}=(2,1)$ for even $i$. This means that the first stock alternatively doubles and halves, while the second stock remains constant. Suppose we choose the CRP strategy which invests half its earnings in each stock at every time period, so $b=(\frac{1}{2},\frac{1}{2})$. Then for odd $i$, our earnings are multiplied by $\frac{1}{2}\times\frac{1}{2}+\frac{1}{2}\times 1=\frac{3}{4}$, and for even $i$ our earnings are multiplied by $\frac{1}{2}\times 2+\frac{1}{2}\times 1=\frac{3}{2}$. Therefore every two timesteps our earnings increase by a factor of $\frac{9}{8}$, showing how CRPs can take advantage of volatility in stock prices. \textbf{Given} a sequence $\{x_i\}_{i=1}^T$, an optimal CRP achieves wealth $S_{T}^{*}:=\max_{b \in \Delta_{m}} S_{T}(b)$. Here are some other useful properties of the best CRP which are trivial to verify:

\begin{proposition} (From [Cover91]): We have the following inequalities:

$$S_T^{*} \geq \max_{j=1,...,m} S_T(e_j)$$

(Best CRP exceeds best stock)

$$S_T^{*} \geq \left (\prod_{j=1}^m S_T(e_j)  \right)^{\frac{1}{m}}$$

(Best CRP exceeds value line)

$$S_T^{*} \geq \sum_{j=1}^m \alpha_jS_{T}(e_j)$$

(Best CRP exceeds arithmetic mean) where $(\alpha_1,...,\alpha_m) \in \Delta^m$.
\

Additionally, the earnings achieved by a CRP is invariant under the permutations of the sequence $\{x_i\}_{i=1}^T$.

\end{proposition}

In general, we will be interested in minimizing the "regret" 

$$LR_{T}:=\ln(S_{T}^{*})-\ln(S_{T})$$

a trading strategy is considered "universal" if, as a function of $T$ we have

$$\frac{1}{T} LR_{T} \in o(1)$$

the reason for considering the difference in logarithms is that usually the CRP will achieve exponential growth on the market. Thus, if for example $S_{T}^{*}$ grows like $e^{\alpha t}$, then a universal portfolio will have growth at least $e^{\alpha t-o(t)}$, so we get the same growth to first order in the exponent. 

\subsection{Overview of the following sections}

Section 2 will cover Cover's approach to solving the above problem. The remainder of the survey will attempt to collect together a number of advances and extensions to Cover's algorithm, including, \textbf{analysis with transaction costs} (Section 3), \textbf{efficient implementation} (Section 4), and \textbf{adding side information} (Section 5). \textbf{More modern approaches to portfolio regret minimization} will be covered in Section 6.

\section{Cover's Algorithm}

Consider the following algorithm: Set $\hat{b}_{1}=(\frac{1}{m},...,\frac{1}{m})$. For $k+1>1$, set 

$$\hat{b}_{k+1}:=\frac{\int_{\Delta_{m}} b S_{k}(b) db}{\int_{\Delta_{m}} S_{k}(b) db}$$

i.e., $\hat{b}_{k}$ can be interpreted as the weighted average of each $b \in \Delta_{m}$, where the weight of $b$ is equal to the earnings it achieves on the observed sequence $\{x_{i}\}_{i=1}^k$. We will denote $\hat{S}_{k}:=\prod_{i=1}^k \langle \hat{b}_i,x_i \rangle$. 

\begin{theorem} ([Cover91]): We have

$$\hat{S}_T =\frac{\int_{\Delta_{m}} S_{T}(b) db}{\int_{\Delta_{m}} db}=E_{b \sim Unif(\Delta_m)}[S_{T}(b)]$$

$$\hat{S}_T \geq \left (\prod_{j=1}^m S_{T}(e_j) \right)^{\frac{1}{m}}$$

and that $\hat{S}_T$ is invariant under permutations of the sequence $\{x\}_{i=1}^T$

\end{theorem}

The first claim says that $\hat{S}_T$ is equal to the average performance on $\{x_i\}_{i=1}^T$ of a randomly chosen CRP. The second claim follows by applying Jenson's inequality to the first equality framed as an expectation. The third claim is a direct consequence of the first. 

\begin{theorem} ([Cover91]): We achieve

$$\max_{\{x_{i}\}_{i=1}^T }\frac{1}{T}\left[\ln(S_{T}^{*})-\ln(\hat{S}_{T})\right]=o(1)$$

i.e. this is a universal trading strategy. 

\end{theorem}

Cover's original paper gives a relatively lengthily proof of theorem 3; in contrast, Blum and Kalai give a more concise proof of Theorem 3 in [BK99] which generalizes Cover's bounds to handle transaction costs, which we cover in the next section. 

\section{Incorporating transaction costs}

In our current model, we have assumed that we can trade without any transaction costs. This assumption is examined in [BK99], and a simple extension to Cover's algorithm is given to handle these extra costs. In particular, consider a fixed percentage commission $0\leq c<1$. This means that if an investor buys $\$1$ of stock, they must pay $c$ dollars in commission. It is also possible to include a commission for selling stock, however the method for this extension is similar and therefore omitted for clarity. Now, at each timestep, if the previous investment portfolio was $b_{i-1}$ and the new investment portfolio is $b_{i}$, the investor must pay (by selling stock) the minimum amount necessary to rebalance their portfolio from $b_{i-1}$ to $b_{i}$. Computing the optimum rebalancing step can be done efficiently; the question is how to modify Cover's algorithm so that transaction costs are taken into account. Intuitively, as our transaction costs become greater, our trading strategy should react by decreasing the change in our portfolio at each timestep. First, consider the following proof technique, which gives an intuition into exactly why Cover's algorithm works:

\

\begin{theorem} (Theorem 1. in [BK99]): We have

$$\frac{\hat{S}_T}{S^{*}_{T}} \geq \frac{1}{(T+1)^{m+1}}$$

$$LR_{T} \leq (m+1)\log(T+1)$$

\end{theorem}
Here's a summarized form of the proof:
\begin{proof}

Let $b^{*}\in \mathbb{R}^m$ denote an optimal CRP vector for $\{x_i\}_{i=1}^T$. Now for $\alpha =1-\min_{j \in [m]} \frac{b_j}{b^*_j}$ and $b,z \in \Delta_m$, we can write $b=(1-\alpha)b^{*}+\alpha z$. Notice that $b$ invests $\alpha$ of its portfolio in $b^{*}$, which implies the inequality

$$S_T(b) \geq (1-\alpha)^{T} S_T(b^{*})=(1-\alpha)^{T} S_T^*$$

Now, viewing $b$ as a random variable drawn uniformally from $\Delta_m$, and $\alpha$ as being the random variable $1-\min_{j \in [m]} \frac{b_j}{b^*_j}$, we have the identity $\hat{S}_T=E_{b}[S_T(b)]$, and hence

$$\frac{\hat{S}_T}{S^{*}_{T}}=\frac{E_{b}[S_T(b)]}{S_{T}(b^{*})} \geq E_{b}[(1-\alpha)^T]=\int_{0}^1P_{b}[(1-\alpha)^T\geq x]dx$$

Now $P_{b}[(1-\alpha)^T\geq x]=P_{b}[\alpha \leq 1-x^{\frac{1}{T}}]$. This is the probability that a randomly chosen $b' \in \Delta_m$ is close enough to $\beta^{*}$ in the sense that it can be written as $b'= (1-\alpha)b^{*}+\alpha z$ for $\alpha \leq 1-x^{\frac{1}{T}}, z \in \Delta_m$. We can compute this probability by noting that

$$Vol_{m-1}(\{(1-\alpha)b^{*}+\alpha z: z \in \Delta_m\})=Vol_{m-1}(\{\alpha z: z \in \Delta_m\})=\alpha^{m-1}Vol(\Delta_{m})$$

Thus

$$\frac{\hat{S}_T}{S^{*}_{T}} \geq \int_{0}^1P_{b}[(1-\alpha)^T\geq x]dx =\int_{0}^1 \left(1-x^{\frac{1}{T}}\right)^{m-1} dx$$

$$=\frac{1}{\binom{T+m-1}{m-1}} \geq \frac{1}{(T+1)^{m+1}}$$

where the last equality follows from standard calculus. 

\end{proof}

The above proof shows that the reason Cover's algorithm works is because the average performance of a CRP on $\{x_i\}_{i=1}^T$ is dominated by the performance of CRP strategies in a small region around $b^{*}$. The proof not only proves theorem 3, but is also easily extended to give bounds in the case of transaction costs. To do this, we generalize Cover's algorithm by defining $S_{k}^{c}(b)$ to be the earnings achieved with CRP $b$ on $\{x_{i}\}_{i=1}^k$ \textbf{including} the commission $c$ incurred for optimally balancing. Then $\hat{b}_{k+1}^c:=\frac{\int_{\Delta_{m}} b S_{k}^c(b) db}{\int_{\Delta_{m}} S(b)^c db}$, and we compare the performance of our algorithm to $\max_{b \in \Delta_m} S^c_{T}(b)$. Now there are only two changes to the above proof:

\begin{enumerate}

\item The first change is that we have $\hat{S}^c_T \geq E_{b}[S^c_{T}(b)]$. Briefly, this is because when there are no transaction costs, Cover's algorithm keeps a portfolio equal to the average portfolio (weighted by wealth) of every CRP strategy running "in parallel". When we include transaction costs, $E_{b}[S^c_{T}(b)]$ is equal to this "parallel" portfolio expectation with transition costs, but Cover's modified algorithm might be able to perform better than this: this is because it can lower re-balancing commissions by "trading" between portfolios running in parallel. For example, if one CRP wants to buy more of stock 1, then instead of paying a commission, we can transfer some of stock 1 to the CRP from another CRP running in parallel which wants to sell stock 1; more details are provided in the original paper.  

\item If $\forall j \in [m]$ we have $b_j \geq (1-\alpha)b_j^{*}$, then $\frac{\text{single-period profit of $CRP_{b}$}}{\text{single period profit of $CRP_{b^{*}}$}} \geq (1-\alpha)(1-c\alpha)$. The $(1-\alpha)$ term is the same term which appeared in previous analysis. The $(1-c\alpha)$ appears because of transaction costs: After an investment period, we can imagine both portfolios re-balancing (and paying commission) to $b^{*}$. Then $CRP_{b}$ needs to rebalance from $b^{*}$ to $b$, for which it incurs at most an additional factor cost of $c\alpha$. Using $1-c\alpha \geq (1-\alpha)^c$, we therefore get that $S^c_T(b) \geq (1-\alpha)^{(1+c)T} S^c_T(b^{*})$. 

\end{enumerate}

Applying these two modifications results in

\begin{theorem} (Theorem 2 in [BK99]) For $0 \leq c \leq 1$

$$\frac{\hat{S}^c_T}{(S^{*})^c_{T}} \geq \frac{1}{((1+c)T+1)^{m+1}}$$

\end{theorem}

\section{Computing Universal Portfolios}

How might one compute $\hat{b}_i$ or $b^{*}$? Numerical integration is too costly, because the time complexity scales exponentially with the number of stocks $m$. [BK99] suggests (given the probabilistic interpretation in section 3) trying a sampling technique of picking uniformally random CRPs $b_1,...,b_k$, equally dividing the initial investment between these CRPs, and then simply running each individual CRP independently on the stock market data already seen ($\{x_{j}\}_{j=1}^{i-1}$) to get an estimate for $\hat{b}_i$. However, the analysis of section 3 shows that the volume of well-performing CRPs in $\Delta_m$ can scale like $T^{-m+1}$ in the worst case; in other words, one would need exponentially many samples in $m$ in order to have non-vanishing probability of picking at least one random CRP which achieves good competitive performance. 

\

Despite this observation, it turns out that random sampling was a valid direction to follow: more than 20 years after Cover published his paper on Universal Portfolios, [AS02] showed that that one can efficiently obtain an approximation to $\hat{b}_i$ through a random walk procedure which samples from the density $ds_{k}(b)=\frac{S_{k}(b) db}{\int_{\Delta_m} S_{k}(b) db}$. The details of the procedure can be found in the original paper, however the key observation which enables efficient sampling is the fact that $\log S_k(b)$ is concave. This allowed the authors to apply a prior result by Frieze and Kannan [FK99] on sampling from log-concave distributions to obtain the following result:

\

\begin{theorem} (Theorem 2 in [AS02]): There is a sampling algorithm and a constant $A$ such that $\forall \epsilon, \eta, \delta, \delta_0$ with

$$\delta_0 \leq \frac{\epsilon}{8mT(m+T^2)}$$

$$\delta \log(\frac{1}{\delta})=\frac{\epsilon \delta_0}{A(m+T)^2}$$

the algorithm generates $64T^2(m+T)\ln(\frac{mT}{\eta})/\epsilon^2$ samples with $\frac{An}{\delta^2}\log(\frac{m+T}{\epsilon \delta})$ random walk steps per sample, and achieves earnings at least $(1-\epsilon)$ times that of Cover's universal algorithm with probability at least $1-\eta$. 

\end{theorem}

Thus there exists an algorithm which will get us within a constant factor of Cover's algorithm with high probability in $poly(T,m)$ time. This result cannot directly be applied to universal portfolios with transaction costs, because the rebalancing step might cause $S_k^c(b)$ to not be log-concave. However if one were to be content with a naive rebalancing strategy (as opposed to rebalancing optimally at each step), one may be able to prove that this rebalancing strategy leads to $S_k^c(b)$ being log-concave, and then adapting the analysis from Section 3 to give a performance guarantee.

\section{Including Side Information}

While Cover's algorithm makes almost no assumptions on $\{x_i\}_{i=1}^T$, realistically one might have extra information about $\{x_i\}_{i=1}^T$ which could help predict future stock prices. At time $i$, this extra information could range from random noise sampled from a pseudorandom number generator, to the weather forecast a week from now, to the actual value of $x_{i+1}$. In other words, the relationship between the side information and $\{x_i\}_{i=1}^T$ could be arbitrary.

\

To formalize this concept, we introduce a side information sequence $\{y_i\}_{i=1}^T$ where each $y_i$ takes a value in $[k]$. At time $i \in [T]$, our algorithm receives $y_i$ as input before choosing a portfolio $b_i$. Let $I_{j}:=\{i \in [T]| y_{i}=j\}$ be the set of times at which $y_i=j$. We define $S^j_{T}(b)=\prod_{i \in I_{j}} \langle b,x_{i} \rangle$, and likewise $\hat{b}^j_i:=\frac{\int_{\Delta_m} b S^j_i(b) db}{\int_{\Delta_m} db}$. When we observe $y_i=j$ at time $i$, Cover's algorithm will choose portfolio $\hat{b}^j_i$. Another way of understanding this approach is to let $(\{x_i\}_{i=1}^T)|_{y_i=j}$ be the sequence $\{x_i\}_{i=1}^T$ restricted to $y_i=j$, i.e. $x_i \in (\{x_i\}_{i=1}^T)|_{y_i=j}$ iff $y_i=j$. Essentially, we partition $\{x_i\}_{i=1}^T$ according to the values of $\{y_i\}_{i=1}^T$, and then we run an independent copy of Cover's algorithm on each partition $(\{x_i\}_{i=1}^T)|_{y_i=1},...,(\{x_i\}_{i=1}^T)|_{y_i=k}$. It follows that the resulting earnings achieved is $\hat{S}_T=\prod_{j=1}^k \int_{\Delta_m} S^j_T(b) db$. In a similar spirit, we define the earnings of the best CRP in hindsight as $S^{*}_T=\sup_{b^1,...,b^k} \prod_{j=1}^k S^j_T(b^j)$, i.e. we are allowed to have a different CRP strategy for each of the $k$ partitions of $\{x_i\}_{i=1}^T$ determined by $\{y_i\}_{i=1}^T$. 

\begin{theorem} (From [CO96]): 

$$\sup_{\{x_i\}_{i=1}^T, \{y_i\}_{i=1}^T } \frac{1}{T}\left[ \log(S^{*}_T)-\log(\hat{S}_T) \right] \leq \frac{k(m-1)}{T}\log(T+1) \rightarrow 0$$

as $T \rightarrow \infty$.

\end{theorem}

If we understand the above algorithm as essentially running $k$ copies of Cover's algorithm on each partition of $\{x_i\}_{i=1}^T$, then theorem 7 follows (up to logarithmic terms) immediately from theorem 4. 

\

The reader may be dissatisfied with this result; not only do we need our side information to come from a finite set, but it also seems that we haven't added anything new to Cover's algorithm. Under no assumptions about the relation between side information and $\{x_i\}_{i=1}^T$, it is unsurprising that $y_i$ must come from a finite set: for example, if $y_i \in \mathbb{R}$ and each $y_i$ were distinct, then there is really no way of "learning" from experience what each $y_i$ means.  

\

One way we can circumvent this problem is to is consider $y_i \in \mathbb{R}^d$ (other metric spaces can be used), and we make the assumption that if $y_i$ and $y_j$ are close in Euclidean norm, then they give "similar" information about the future. Hazan and Megiddo take this approach in [HM07] and prove a regret bound with side information in a more general setting of \textbf{online convex regret minimization}: we imaging that at each time $i \in [T]$ we receive a convex function $f_i$, and then we receive a cost $f_i(b_i)$. In the current setting of this survey, we would pick $f_i(b_i):=-\log(\langle b_i,x_i \rangle )$, however the authors note that one can get better regret bounds for this specific choice of $f_i$ (see the result in the next section for online exp-concave regret minimization). For this problem, the following result can be achieved:

\

\begin{theorem} (Simplified version of Theorem 4 in [HM07]):  Let $W$ denote the maximum possible distance between any $y_i,y_j$. Suppose we restrict each $b_i$ to come from a convex set $F \subset \Delta_m$, and let $G:=\sup_{b \in F, i \in [T]}\|\nabla f_i(b)\|$. Let $X_L$ denote the family of mappings from $\mathbb{R}^d$ to $F$ which have Lipschitz-constant $L$. Then there exists an algorithm which when given side information $\{y_i\}_{i=1}^T$ achieves regret

$$\sum_{i=1}^T f_i(b_i)-\sup_{v \in X_L} \sum_{i=1}^T f_i(v(y_i)) \leq O(WGL^{1-\frac{2}{d+2}}T^{1-\frac{1}{d+2}})$$

moreover, this algorithm runs in time polynomial in T,m and d. 
\end{theorem}

Notice that this algorithm gives a universal trading strategy if $W$ is bounded. The details of the algorithm can be found in the original paper, but at a high level the algorithm works as follows: We construct an $\epsilon$-Net on the set of side-information points we receive. Recall that an $\epsilon-$Net around a set of points $S$ is a set of points $N$ such that for every $x \in S$, the minimum distance between $x$ and a point in $N$ is $\leq \epsilon$, and for any $x,y \in N$, we have $\|x-y\| \geq \epsilon$. The idea is that if $y_i$ and $y_j$ are sufficiently close, then $v(y_i)$ and $v(y_j)$ must also be close, and so we don't lose much performance by approximating a new piece of side information $y_i$ by its nearest neighbour $\tilde{y}_i$ in the $\epsilon$-Net. In this way, each point in the $\epsilon$-Net is a "representative" for nearby side-information points. For each fixed representative $\tilde{y}$, the algorithm then partitions $\{x_i\}_{i=1}^T$ according to these representatives, and then runs a general online regret minimization gradient descent method (detailed in [Zin03]) on each partition. Essentially, this is almost the same as Cover's algorithm with side information: In Cover's algorithm, we ran an independent copy of a regret minimization algorithm on $\{x_i\}_{i=1}^T$ partitioned by the values in $\{y_i\}_{i=1}^T$, the only difference is that we partition $\{x_i\}_{i=1}^T$ by the values in $\{\tilde{y}_i\}_{i=1}^T$ instead. The regret bounds are derived by summing the regret over all points in the $\epsilon$-Net, and showing that using a representative $\tilde{y}_i$ doesn't impact performance significantly. 

\

We therefore see a general approach that exists for regret minimization with side information: run independent copies of a regret minimization algorithm without side information, where each copy only deals with time periods associated to a particular piece of side information. If we make assumptions on how different values of side information $y_i$ are related, we can partition the space of side-information values and have each regret minimization algorithm work on corresponding regions of side information points.

\section{Other approaches to portfolio regret minimization}

\subsection{Multiplicative updates}

So far, we have only considered variants of Cover's algorithm for achieving Universal trading algorithms. Cover's paper was published in 1991; In 1998 (before it was known how to run Cover's algorithm in polynomial time), Helmbold et al. showed in [HSSW98] that a far simpler and more efficient strategy could be used to achieve a Universal portfolio, albeit with worse regret bounds than Cover's algorithm. In particular, consider the function defined for $b_{i+1} \in \Delta_{m}$:

$$F(b_{i+1})= \eta \log(\langle b_{i+1},x_{i} \rangle)-d(b_{i+1},b_{i})$$

where $d$ is a distance function, which in this case is taken to be the KL divergence between $b_{i+1}$ and $b_{i}$, i.e. $d(u,v):=\sum_{j=1}^m u_j \log(\frac{u_j}{v_j})$. [HSSW98] proposes an algorithm which chooses $b_{i+1}$ by approximately maximizing $F(b_{i+1})$. In particular, if we replace the first $\log$ term in $F$ with a first order Taylor series approximation around $\langle b_{i}, x_{i} \rangle$, then maximizing this approximation is equivalent to the update

$$(b_{i+1})_{j}=\frac{(b_{i})_{j}\exp(\eta (x_{i})_{j}/\langle b_{i},x_{i} \rangle)}{\sum_{k=1}^m (b_{i})_{k}\exp(\eta (x_{i})_{k}/\langle b_{i},x_{i} \rangle)}$$

leading to a very efficient update rule. Using this update rule, one is also able to get the following theorem: 

\begin{theorem} (Theorem 4.1 in [HSSW98]): Let $u \in \Delta_{m}$ be a portfolio vector, and suppose $\forall j \in [m], \forall i \in [T],$ we have $(x_{i})_{j} \geq r >0$, and $\max_{j}(x_{i})_{j}=1$. Then for $\eta >0$, we have

$$\sum_{i=1}^T \log(\langle b_{i},
x_{i}\rangle)  \geq \sum_{i=1}^{T} \log(\langle u,
x_{i}\rangle)-d(u,b_{1})/\eta-\frac{\eta T}{8r^2}$$

If we take $b_1$ to be the uniform vector, and $\eta=2r\sqrt{2\log(M)/T}$, then this bound becomes

$$\sum_{i=1}^T \log(\langle b_{i},
x_{i}\rangle)  \geq \sum_{i=1}^{T} \log(\langle u,
x_{i}\rangle)-\frac{\sqrt{2T\log(M)}}{2r}$$

\end{theorem}

Notice that the regret has a $T^{\frac{1}{2}}$ dependence instead of the $\log(T)$ dependence in Cover's algorithm; however, the dependence in the number of stocks is only {\em logarithmic}. The assumption $\forall i, \max_{j}(x_{i})_{j}=1$ is not restrictive, because the logarithmic regret $LR_T=\sum_{i=1}^{T} \log(\langle u,
x_{i} \rangle)- \log(\langle b_{i},
x_{i} \rangle)$ is invariant under rescaling any particular $x_i$ by a positive constant. The assumption $(x_{i})_{j} \geq r >0$ can be removed by slightly modifying the choice of $b_{i}$, and leads to a universal portfolio algorithm. Theorem 9 is proven by bounding the difference $d(u,b_{i+1})-d(u,b_{i})$ by $-\eta \log(\langle u, x_{i} \rangle /\langle b_{i} , x_{i} \rangle)+\frac{\eta^2}{8\langle b_{i},x_{i} \rangle^2}$. Using $-d(u,b_{1})\leq d(u,b_{T})-d(u,b_{1})$ and $d(u,b_{T})-d(u,b_{1})=\sum_{i=1}^T d(u,b_{i+1})-d(u,b_{i})$ then gives the theorem result. A key part of this analysis is the choice of $d$ as the KL divergence, which conveniently allows us to get bounds in terms of the regret $LR_{T}$. 

\subsection{Newton steps and follow the leader}

Around 2005, a new analysis technique was proposed by Agarwal and Hazan which allowed the construction of algorithms which provably achieved optimal regret (equal to the regret of Cover's algorithm), while also being computationally efficient. These algorithms implement a "follow-the-best-expert" strategy using Newton steps to find the best expert. [AHKS06] covers one particular Newton step strategy to achieve $LR_{T}=\log(T)$ under the assumption that $(x_i)_{j}$ is bounded away from $0$, and $LR=\sqrt{T}$ without this assumption. 

\

In [HK12], Hazan and Kale give a relatively simple algorithm for online exp-concave regret minimization (which includes the setting of this survey as a special case) with an interesting regret bound of $LR_{T} \leq O(\log(Q))$ (but linear in $m$), where $Q$ is the variation of the price sequence $\{x_i\}_{i=1}^T$, appropriately normalized so that $\forall i \in [T], \max_{j} (x_i)_j = 1$. This bound is interesting for three reasons: first, notice that this bound reduces to Cover's regret bound of $O(\log(T))$ in the worst case. Secondly, this bound is independent of the number of trading time periods: if an investor wants to trade more frequently over a 30 day period, Cover's bounds automatically deteriorate, while the bound of [HK12] will perform better provided that price vectors have lower volatility over shorter time periods (which can be a reasonable assumption). Lastly, it is empirically true that stocks can sometimes loosely behave as random walks with low variance, in which case the regret bound of $O(\log(Q))$ becomes particularly strong. The algorithm which achieves these regret bounds is simply to "follow the regularized leader":

$$b_{i}:=\arg \min_{b \in \Delta_m} \left (\sum_{k=1}^{i-1} -\log(\langle b , x_{k}\rangle)+\frac{1}{2} \|b\|^2 \right)$$

where each $x_i$ has been scaled so that $\max_{j} (x_i)_j=1$. The analysis of the performance of this algorithm is lengthy and relies on a number of inequalities, but the method essentially boils down to the following: (1) using a stability lemma from [KV05], bound the regret of this follow the leader algorithm by the difference  $\sum_{i=1}^T -\log(\langle p_t,x_t \rangle)+\log(\langle p_{t+1},x_t \rangle)+\text{(some regularization term)}$. (2), use convexity of $\log$ to bound this regret by a function involving the derivative of $\log$. (3), bound a function involving the difference in gradients $\nabla \log(\langle p_{t+1},x_t \rangle)-\nabla \log(\langle p_{t},x_t \rangle)$ by a function involving $\nabla^2 \log(\langle p_{t+1},x_t \rangle)$. Overall, the authors claim that some of the analysis is based on second order Newton step ideas from [HAK07]. In addition, one needs to extend the ideas in [HAK07] by bounding the difference between successive predictions $b_{i}$ and $b_{i+1}$ by norms which change during each timestep, which allows the bounds to be sensitive to the variation in $\{x_{i}\}_{i=1}^T$.

\

\section{Conclusion, future directions}

We've seen two key strategies to online portfolio risk minimization: track an average of all possible CRPs, or follow some kind of regularized leader. We've also seen how regret minimization algorithms in this area have been developed over time to deal with transaction costs and side information. A general approach to dealing with side information is given by treating sub-sequences of $\{x_i\}_{i=1}^T$ as independent problems partitioned on their side information content, and then bounding the total regret as a sum of the regret of each subproblem. There are also other extensions to CRP regret minimization which have not been covered, such as "switching portfolios" in [Singer13], where one competes against an expert which can change its portfolio at most $K$ times. Again, some approaches to this "switching CRP" problem involve simply averaging over all switching CRP experts, and then paying a penalty regret factor which is linear in $K$. 

\

After reading this summary of the covered techniques, one might get a sense that the approaches to regret minimization in this area are rather {\em simple}, even if the analysis of their performance is nontrivial. While there are still open problems regarding the best regret bounds we can get in terms of $m$,$T$ and the price variation (see the end of [HK12]), there is an overarching limitation due to the generality of the regret minimization problem we are considering. For example, if we make {\em no assumptions} on the relation between side information and $\{x_i\}_{i=1}^T$, it is unsurprising that the best approaches available reduce to breaking the sequence $\{x_i\}_{i=1}^T$ into independent subproblems (there are also settings where an information theoretic argument will show that this gives the best asymptotic bounds). Likewise, if we make no realistic assumptions on $\{x_i\}_{i=1}^T$, it is perhaps unsurprising that the strategies which give optimal regret (when $m$ is viewed as a constant) involve either averaging experts or following leaders.  

\

There are therefore two further directions one might look at in the area of portfolio regret minimization. The first is to determine trade-offs between regret dependence on $T,m$, and the price variation. For example, we have seen some algorithms with regret logarithmic in $m$, yet polynomial in $T$, or polynomial in $m$ yet logarithmic in $T$. Other algorithms have performance which depends on the value of $\min_{i,j} (x_i)_{j}$, while the performance of Cover's algorithm does not depend on this quantity. A second direction is to try and extend the current framework of this problem to deal with more informative information. In a realistic setting, market analysts have some idea of what the market will do in the future, and so it might be natural to restrict the regret minimization problem only to sequences $\{x_i\}_{i=1}^T$ which analysts deem are plausible. Besides potentially allowing one to prove better regret bounds, solving this problem may give an objective approach to pricing market information. For example, if we have some information $I$ which restricts the set of possible sequences $\{x_i\}_{i=1}^T$, we might say that the {\em value} of the information $I$ is equal to the maximum regret our algorithm could have made before knowing $I$, minus the maximum regret it could make knowing $I$. A trivial example of "valuable" information which applies to some existing algorithms is the value of $r=\min_{i,j} (x_i)_{j}$ or $T$; if we don't know $r$ or $T$ in advance, algorithms can sometimes get around this with techniques such as standard "doubling tricks" used in regret minimization, albeit with slightly worse regret bounds. Thus we might say that {\em knowing} $r$ in advance has positive market value for some algorithms. Once we know how to price information, this opens up the exciting prospect of solving the inverse question: how to find non-trivial information content (restrictions on $\{x_i\}_{i=1}^T$) with high market value.

\bibliographystyle{alpha}

\end{document}